\documentclass{aa}
\usepackage{graphicx}
\usepackage[varg]{txfonts}
\usepackage{natbib}
\usepackage[colorlinks=true, linkcolor=blue, citecolor=blue, filecolor=blue, urlcolor=blue]{hyperref}

\begin{document}

   \title{Low velocity streams inside the planetary nebula H\,2-18}

   \subtitle{A 3D photoionization and kinematical reconstruction\thanks{Based on observations collected at the European Southern Observatory under ESO programme 099.D-0386(A).} }

   \author{K. Gesicki
      \inst{1}
      \and
      A. Zijlstra
      \inst{2,3}
      \and
      M. Hajduk
      \inst{4}
      \and
      A. Iwanowska
      \inst{1}
      \and
      K. Grzesiak
      \inst{1}
      \and
      K. Lisiecki
      \inst{1}
      \and
      J. Lipinski
      \inst{1}
      }

   \institute{Institute of Astronomy, Faculty of Physics, Astronomy and Informatics, Nicolaus Copernicus University,\\ ul. Grudziadzka 5, 87-100 Torun, Poland, \email{kmgesicki@umk.pl}
         \and
             Jodrell Bank Centre for Astrophysics, School of Physics \& Astronomy, University of Manchester,\\ Oxford\,Road, Manchester M13 9PL, UK
        \and
            School of Mathematical and Physical Sciences, Macquarie University, Sydney, NSW 2109, Australia         \and
            Department of Geodesy, Faculty of Geoengineering, University of Warmia and Mazury,\\ ul.\,Oczapowskiego 2, 10-719 Olsztyn, Poland
             }

   \date{Received ; accepted }

% \abstract{}{}{}{}{}
% 5 {} token are mandatory

  \abstract
  {} % leave it empty if necessary
   {Numerous planetary nebulae show complicated inner structures not obviously explained. For one such object we undertake a detailed 3D photoionization and kinematical model analysis for a better understanding of the underlying shaping processes.}
   {We obtained 2D ARGUS/IFU spectroscopy covering the whole nebula in selected, representative emission lines. A 3D photoionization modelling was used to compute images and line profiles. Comparison of the observations with the models was used to fine-tune the model details. This predicts the approximate nebular 3D structure and kinematics.}
   {We found that within a cylindrical outer nebula there is a hidden, very dense, bar-like or cylindrical inner structure. Both features are co-axial and are inclined to the sky by 40 deg. A wide asymmetric one-sided plume attached to one end of the bar is proposed to be a flat structure. All nebular components share the same kinematics, with an isotropic velocity field which monotonically increases with distance from the star before reaching a plateau. The relatively low velocities indicate that the observed shapes do not require particularly energetic processes and there is no indication for the current presence of a jet. The 3D model  reproduces the observed line ratios and the detailed structure of the object significantly better than previous models.}
  {}


   \maketitle
%
%________________________________________________________________

\section{Introduction}

The theory of mass loss at the late asymptotic giant branch evolutionary phase is still in development and this is one of the most important missing factors in the stellar evolution theory. The ejected matter often takes on interesting axi- or a-symmetric shapes and the understanding of physical processes causing these is far from clear. Theoretical approaches need
some suggestions and hints derived from observations.
Such hints can be found from studies of planetary nebulae (PN).

\begin{figure}
   \resizebox{\hsize}{!}{\includegraphics{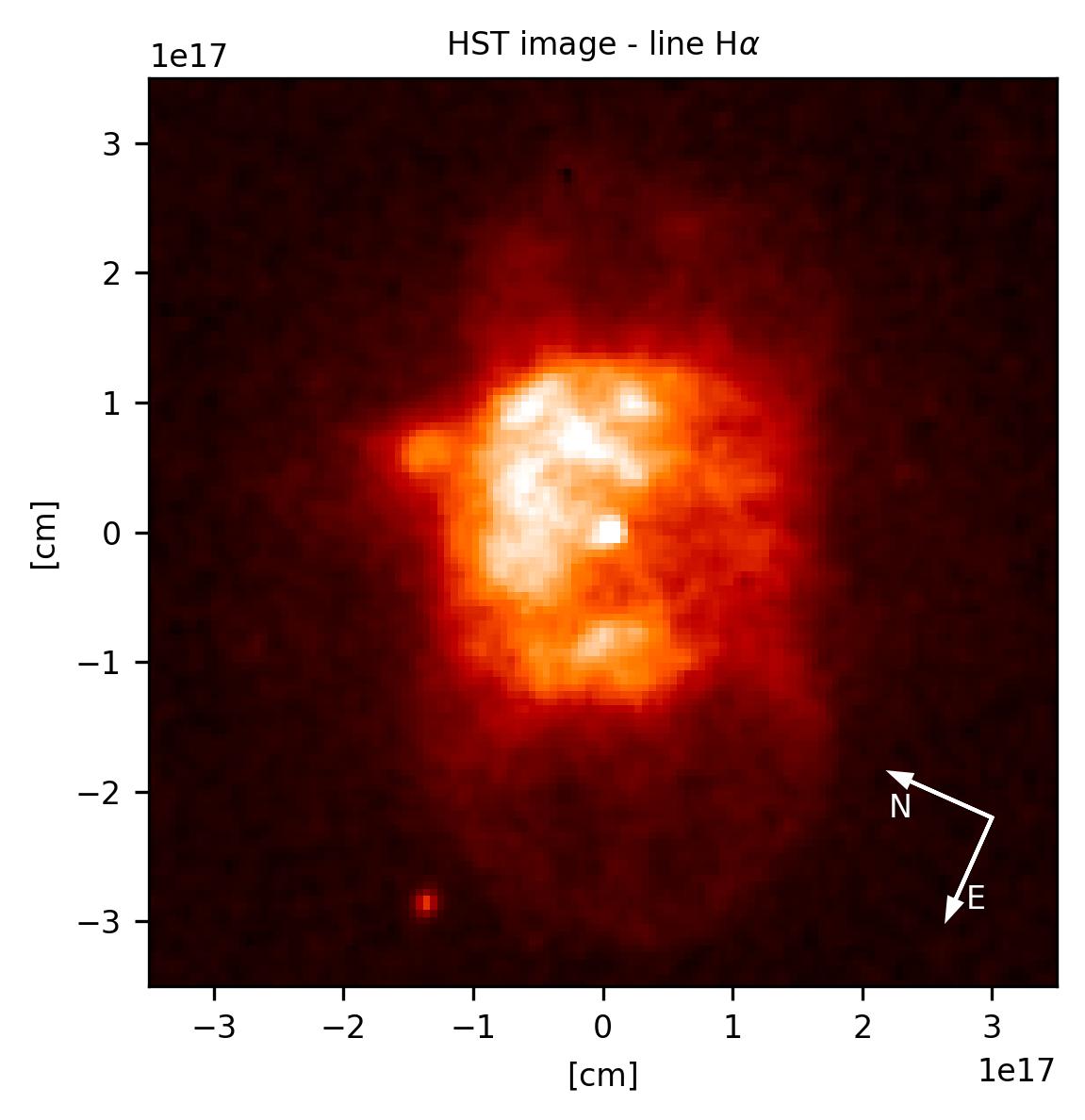}}
   \caption{The HST image of the PN H\,2-18 obtained in an H$\alpha$ filter.
      The plot is in linear intensity scale. The angular dimensions are converted to $10^{17}$\,cm assuming that the PN is at the Galactic Bulge distance of 8\,kpc and centred at the star position. The image is rotated to easier compare with computed structures.}
   \label{HSTima}
\end{figure}

A classical concept of a PN is a
shell of gas ejected by a star at the end of its life.
Initially the shell is opaque and hidden from view.
Only after the shell expands sufficiently for the hot central core to shine through and ionize the nearby material does the object enter the PN phase. This phase ends when the ejecta expand  to a large volume and very low density, or when the remaining stellar core enters the White Dwarf cooling track and fades significantly. Both timescales are comparable.

We study details of PNe by applying the so-called kinematical
reconstruction (sometimes called a `reverse engineering').
This derives the nebular structure and
kinematics from line profiles based on assumed velocity fields which are verified through
photoionization and emission line profiles computation \citep{2000A&A...358.1058G}.  For this
method to work and to reduce the inherent ambiguities, both high
quality images and high resolution spectroscopy are needed in order to constrain the density and velocity spatial distributions.

\begin{figure*}
   \resizebox{\hsize}{!}{\includegraphics{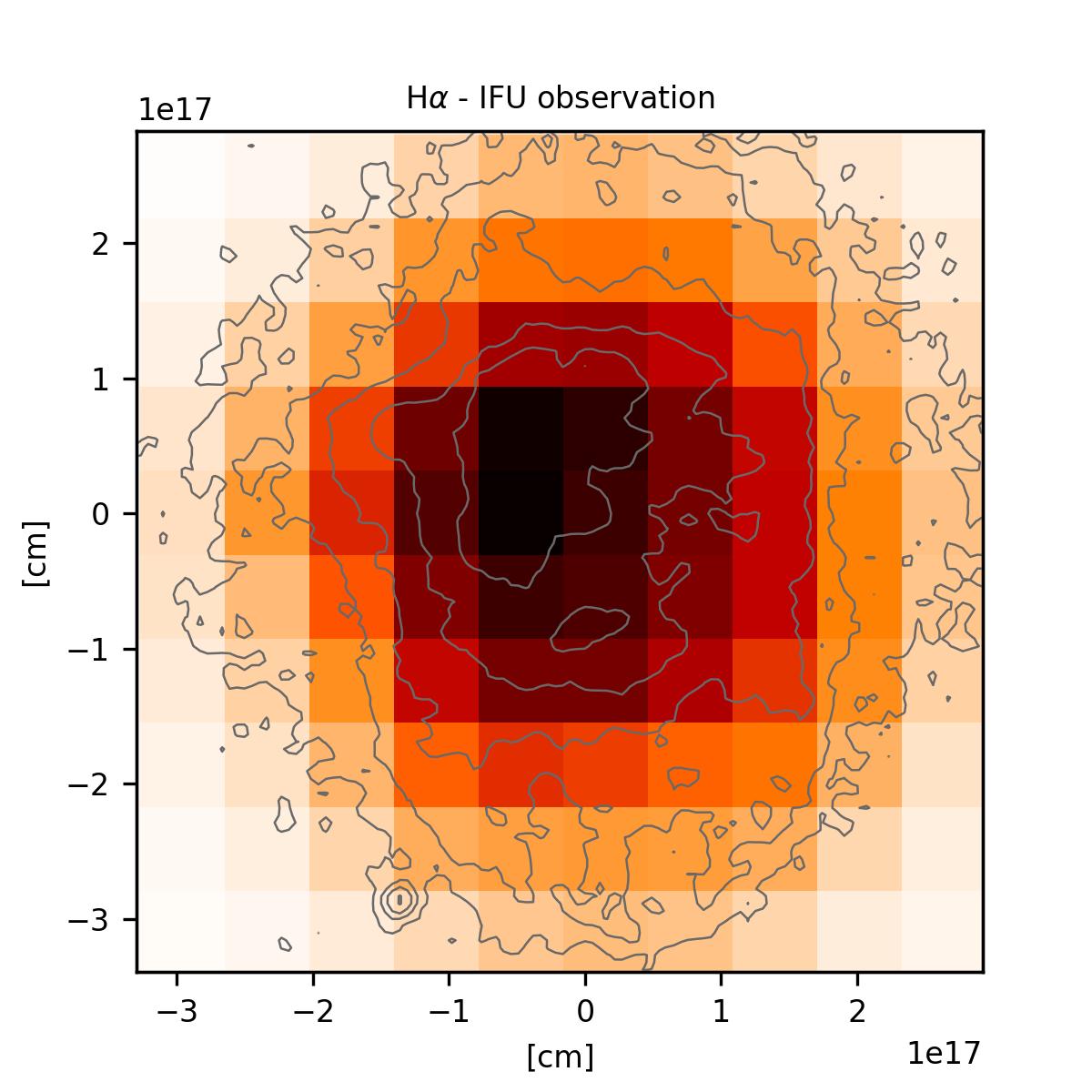}\includegraphics{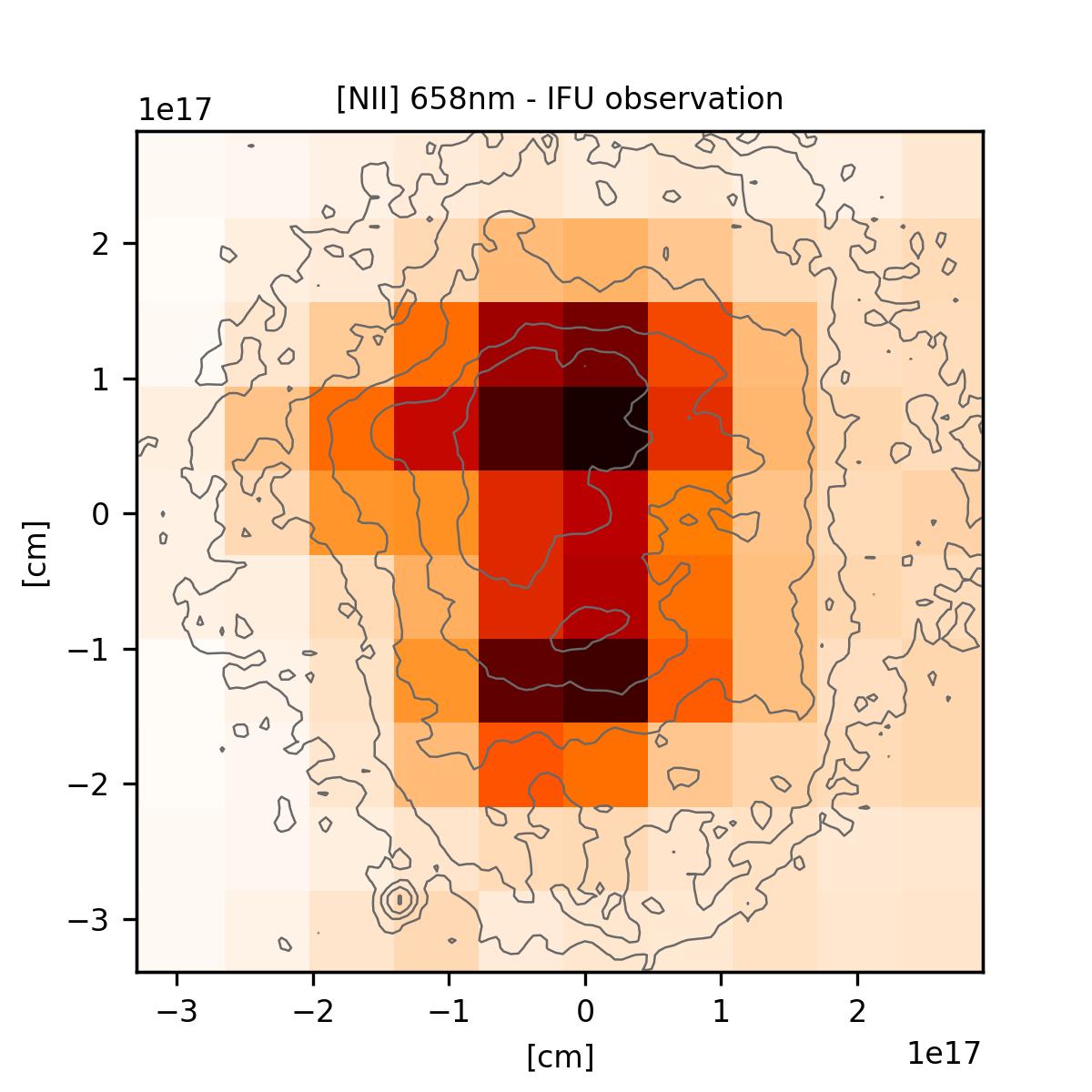}\includegraphics{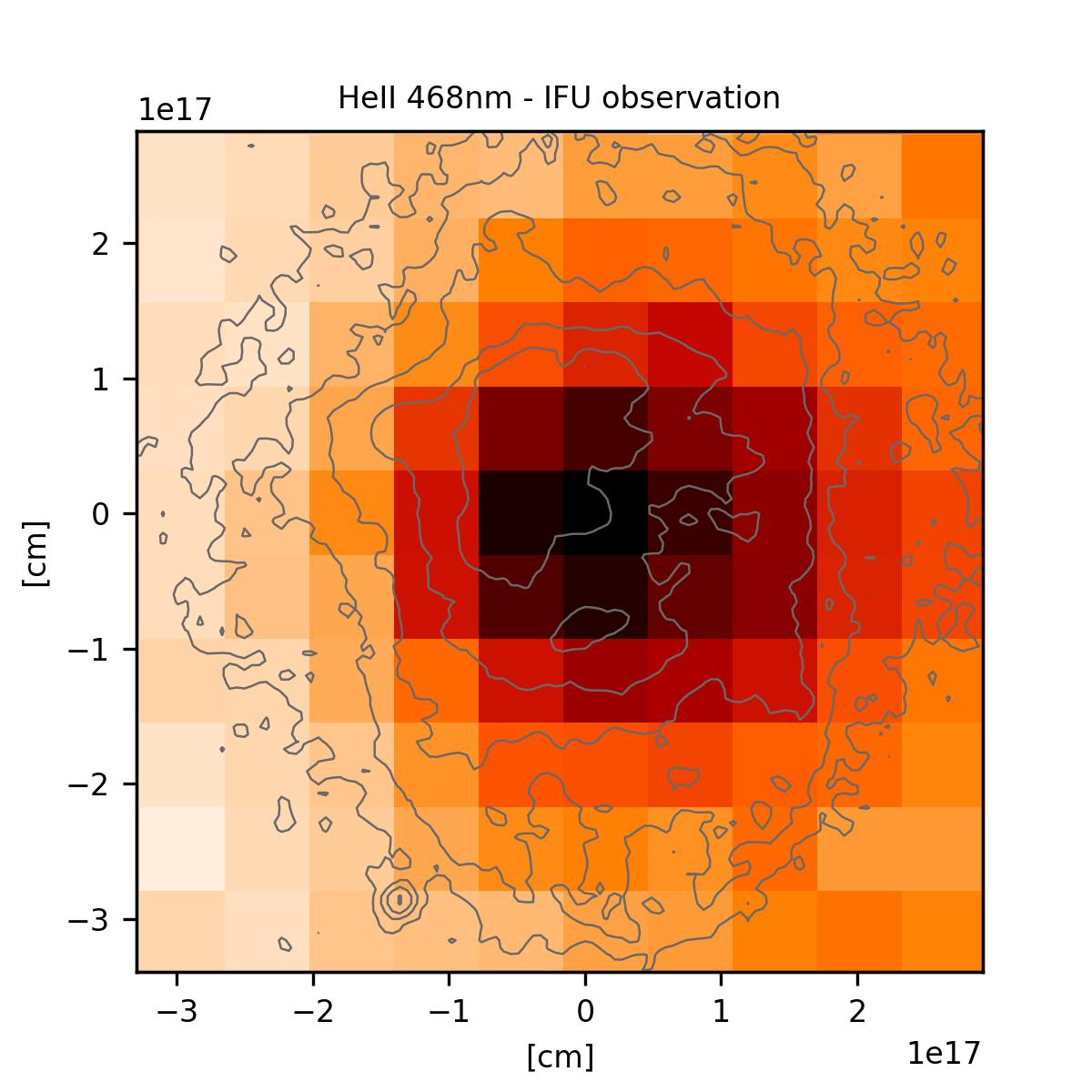}}
   \caption{The ARGUS/IFU monochromatic images with HST contours defined in Fig.\,\ref{mapHa}. The high resolution HST image has been positioned to obtain the best agreement with IFU images of obviously low resolution. Shown are examples of H$\alpha$, [\ion{N}{ii}] 658.3\,nm, and \ion{He}{ii} 468.6\,nm lines.}
   \label{IFUoverplot}
\end{figure*}

Here we present new 2D spectroscopic observations and new 3D
photoionization modelling of a previously analyzed PN, allowing us to compare the new results with the older ones, to verify the new approach and to indicate all its advantages.

The planetary nebula H\,2-18 (see Fig.\ref{HSTima}) was previously analyzed using the kinematical reconstruction approach. \citet{2014A&A...566A..48G} applied the Torun codes and derived a spherically symmetric model of this PN. Later, \citet{2016A&A...585A..69G} used the same observational data but
applied the pseudo-3D code py-Cloudy. Both papers were based
on the HST imaging and ESO/VLT spectroscopic long-slit observations.

In 2017 we obtained ARGUS IFU 2D data of high spatial and spectral resolution of this object. The observations covered selected spectral lines of high and low excitation which probe different nebular regions and allowed for a more precise kinematic and photoionization analysis. The collected data are presented in the Sect.\,\ref{observations}.

For modelling of this nebula we applied the publicly available code MOCASSIN which computes photoionization of a 3D gaseous structure. To initiate the 3D modelling and to compare the model outputs with the 2D spectra we wrote a number of Python scripts. The modelling procedure and the obtained density structure and velocity field are described in the Sect.\,\ref{analysis}.

The nebula H\,2-18 appears to be complex and there are few published studies of similar objects to which it can be compared. The generalization of our unique results is also not easy. Some aspects of this are finally discussed in Sect.\,\ref{discussion}.

%__________________________________________________________________

\section{Observations}
\label{observations}

The ESO FLAMES/ARGUS instrument provided the high-resolution spectroscopy of spatially resolved Galactic Bulge \object{PN H 2-18} (PN G 006.3+04.4; 17:43:28.7, $-$21:09:51.30 (J2000)). The nebula was observed in 2017, on Sep. 18 (wavelengths around 465 and 504\,nm) and on Sep. 21 (627 and 651\,nm). The wide field of view is a mosaic of $22 \times 14$ microlenses each of $0.52 \times 0.52$\,arcsec. The instrument setup was selected to cover at high spectral resolution ($R \gtrsim 30\,000$) the regions around four lines: [\ion{O}{iii}] 500.7\,nm, H$\alpha$ (including [\ion{N}{ii}] 658.3\,nm), \ion{He}{ii} 468.6\,nm and [\ion{O}{i}] 630.2\,nm. For this object at the time of observations, the line [\ion{O}{i}] 630.2\,nm happened to be strongly affected by the overlying terrestrial atmospheric emission which unfavourably overlapped in radial velocity, and so it was excluded from the analysis.

The standard ESO reduction pipeline (EsoReflex) has been applied to the raw data. Further data processing and plotting was performed using Python with modules astropy, scipy and matplotlib. These concern the removal of bad pixels by median filtering the data cube in the XY axes (sky plane) and smoothing the spectral noise by the Savitzky-Golay filter along the wavelength axis. From the original IFU field of $22 \times 14$ pixels we extracted for the analysis a square of $10 \times 10$ pixels which covers the whole nebula and is shown in the figures. The examples of the monochromatic images (i.e. integrated over the full width of a single emission line) are shown in Fig.\,\ref{IFUoverplot}. They show  background emission increasing towards the right edges which was not flat-fielded completely by the pipeline. We did not correct for this because it did not affect the morphological and kinematical modelling. Absolute flux calibration was not attempted, instead the line flux values needed for the model fitting were taken from literature. In all figures we present the observational data with the sky coordinates converted to $10^{17}$ cm (assuming the PN distance of 8\,kpc) and measured with respect to the position of the central star to ease the comparison with models where the real physical size is used. We do not present the [\ion{O}{iii}] 500.7\,nm monochromatic image because it is very similar to H$\alpha$.

Previous HST observations in the H$\alpha$ filter show an  overall axially symmetric, elongated structure which in Fig.\,\ref{HSTima} is rotated so that it is positioned vertically. The most intense, inner broad area is asymmetric with strongest emission on the upper-left of the inner region. Outside of the main cylinder near its middle,  a  low-intensity ears-like extensions can be seen (oriented horizontally in the image) which could be a trace of a faint equatorial ring. 

Our new data confirm these features (Fig.\,\ref{IFUoverplot} left panel) and show more than that. The IFU data revealed an intriguing pair of  small and bright blobs  in the [\ion{N}{ii}] 658.3\,nm image (Fig.\,\ref{IFUoverplot} middle panel); their dominant emission is axially symmetric with a weaker asymmetric component. This feature is completely absent in \ion{He}{ii} 468.6\,nm (Fig.\,\ref{IFUoverplot} right panel) and is barely visible in H$\alpha$. 

We see now that neither a spherical nor an elliptical model is sufficient to describe the PN H\,2-18 and we need to consider the inner component. Although the elongation in the [\ion{N}{ii}] image appears jet-like, because of the high density we estimate  and the low velocity,  we shall rather call it bar-like or cylindrical, leaving the case open.

%__________________________________________________________________

\section{3D model analysis}
\label{analysis}

The photoionization modelling was performed with the publicly available MOCASSIN code \citep{2003MNRAS.340.1136E}. Supplementary codes were written in Python, these concern a construction of a 3D density distribution cube, calculation of the emission line shapes derived from total emissivity with assumed velocity field and emerging from a defined pixel area on the sky, graphical presentation of the observed and modelled emissions, etc. The models presented here were computed on a spatial grid of $121 \times 121 \times 121$ points which ensured sufficient spatial accuracy with reasonable time for simulations and image processing.

We aimed at a model which, while  being as simple as possible, fits the images and simultaneously reproduces the line ratios and velocities.

\subsection{Density structure}

\begin{figure}
   \resizebox{\hsize}{!}{\includegraphics{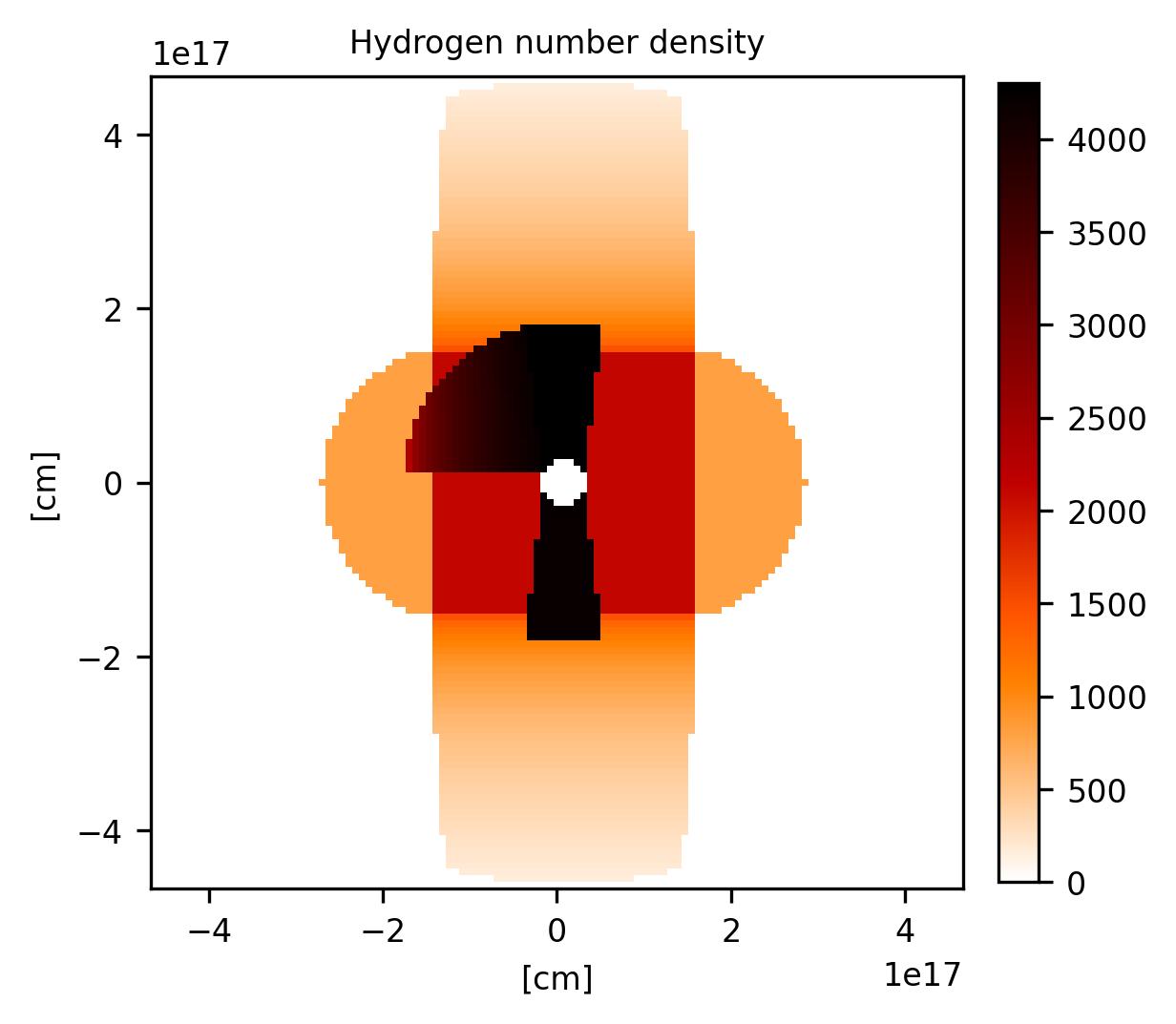}}
   \caption{The density cross-section of the assumed model. The symmetry axis is positioned vertically, all structures (except for the plume) are cylindrically symmetrical. This graph shows the true spatial dimensions of the substructures for a distance of 8\,kpc. The plume is in the plane of the image and extends to $\pm 2.4\times 10^{16}$\,cm above and below it.}
   \label{density}
\end{figure}

\begin{figure*}
   \resizebox{\hsize}{!}{\includegraphics{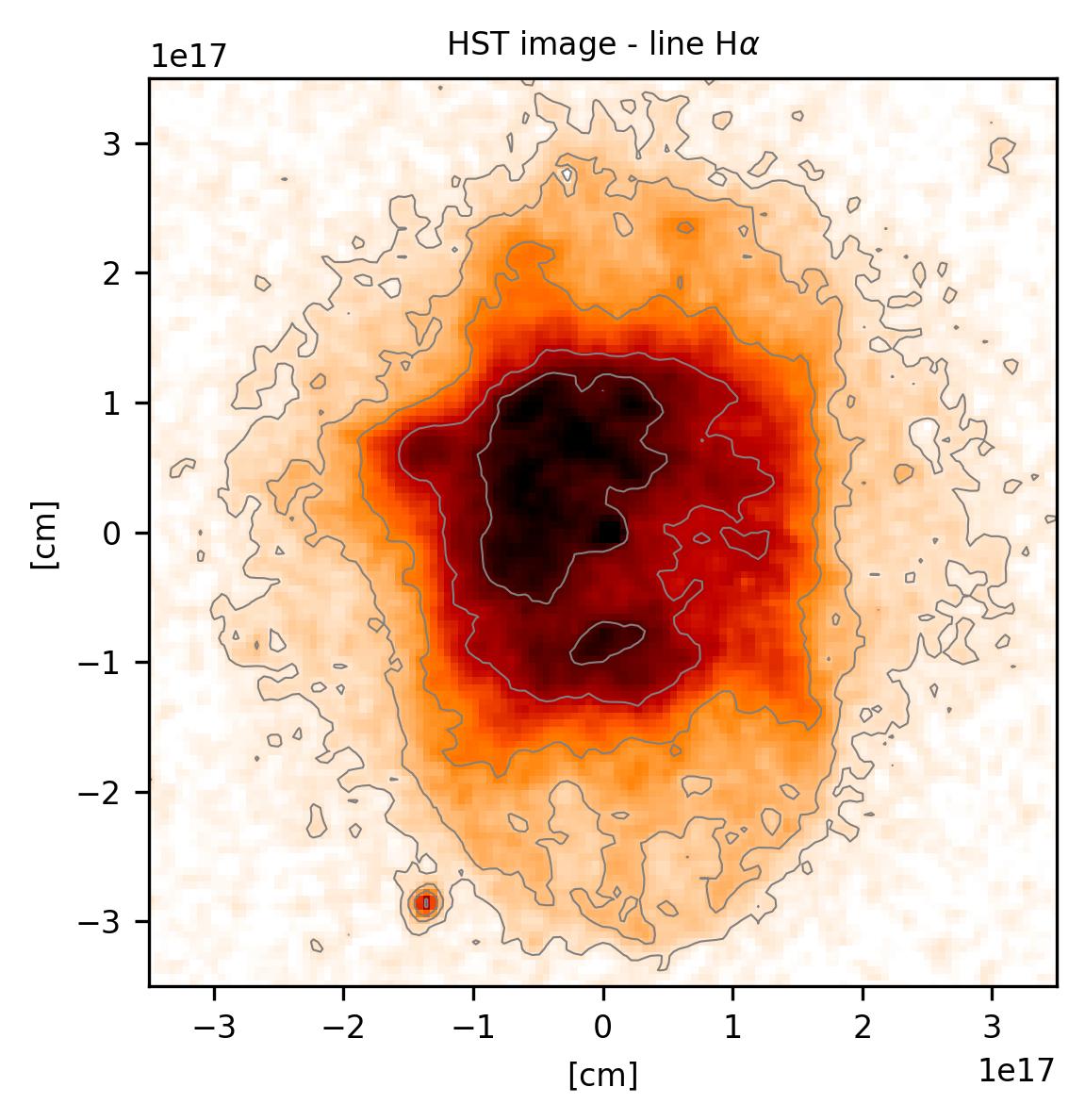}\includegraphics{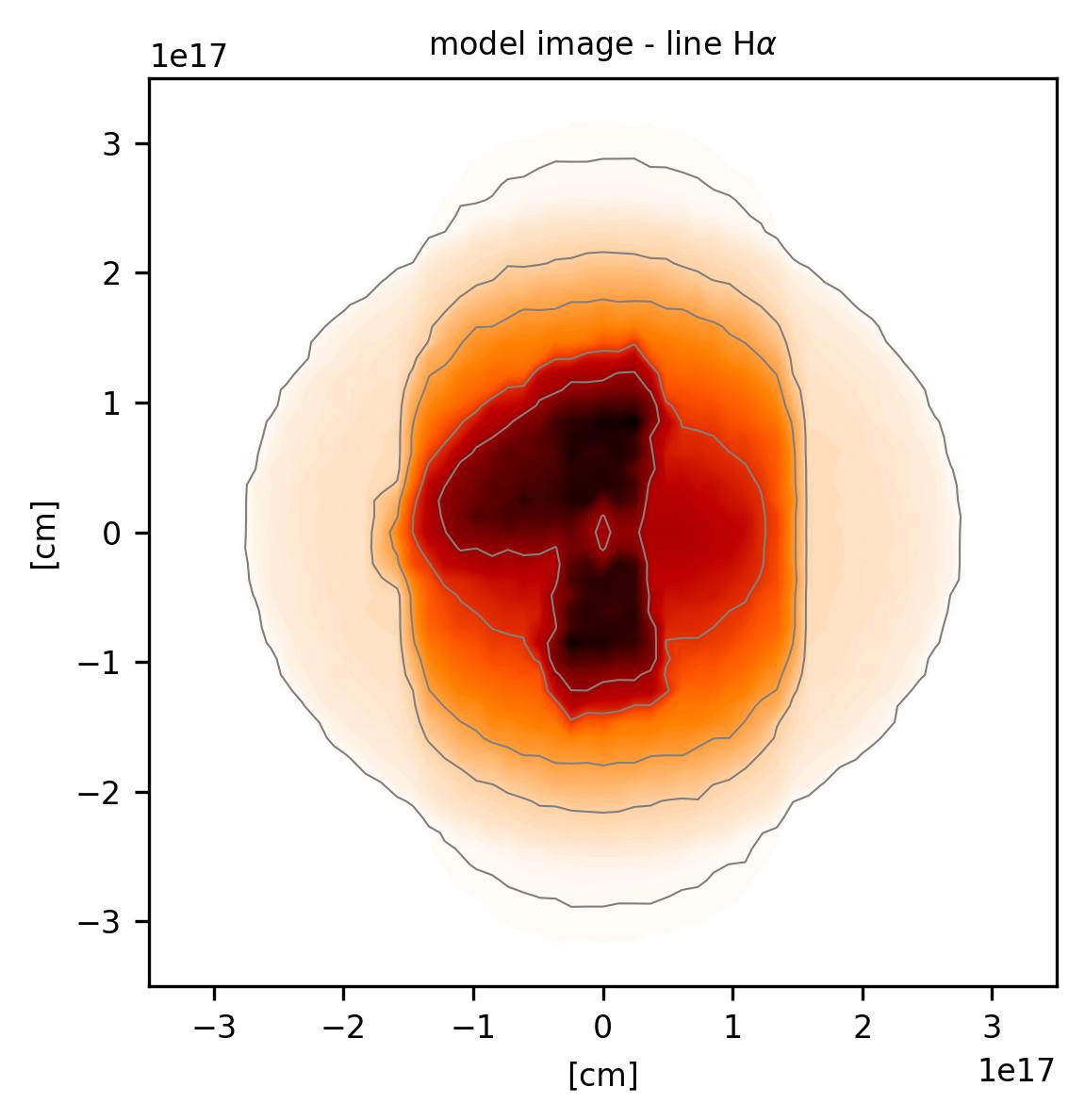}}
   \caption{The observed image in the H$\alpha$ line (left panel) compared with the one obtained from the assumed model (right panel). The axes are the same as in Fig.\,\ref{HSTima}. The contours (0.04, 0.09, 0.19, 0.4, 0.6 of maximum) were selected to bring out the substructures, the same contours are shown in both panels and in all other plots. For the model image the main nebular axis is inclined to the sky plane at 40 deg and because of lack of background noise the lowest contour is set at 0.01 level.}
   \label{mapHa}
\end{figure*}

The components of the model discussed below are presented in Fig.\,\ref{density} as a cross-section of the density distribution taken along the presumed symmetry axis. The values of the hydrogen number density are shown in the side bar, while the size of the nebula in cm is based on assuming the Galactic Bulge distance of 8\,kpc.

The central region within a radius of $2.4\times 10^{16}$\,cm was not considered in the modelling and assumed empty. The ionizing source (central star) was assumed a black-body with temperature $T_\mathrm{bb}$ of 62\,kK (previously it was estimated 52\,kK).

The main nebular body shows nearly straight edges seen clearly in vertical direction in the HST image in Fig.\,\ref{HSTima} (especially as the contours on the right side in the left panel of Fig.\,\ref{mapHa}). This cannot be the ionization boundary because we do not see there the low-excitation [\ion{N}{ii}] 658\,nm emission. To reproduce this feature we assumed a cylindrical outer structure similar to that derived previously in \citet{2016A&A...585A..69G}. The difference concerns the edges which previously were spreading outwards and had higher density. Now we found a reasonable fit assuming a plain cylinder fully filled with gas density decreasing along the axis.

The nebular model structure and inclination was fine-tuned together with the velocity field: both factors provide the emission lines profiles (Sect.\,\ref{velocityfield}) which verify the model. The adopted, enhanced, constant density of the middle part of the main cylinder is responsible for the high intensity of the symmetric component of the H$\alpha$ emission profiles. The asymmetric parts of emission profiles of H$\alpha$ and [\ion{N}{ii}] (see Figs.\,\ref{BarPloHa}, \ref{BarPloN2}) originate in the proposed inner bar and indicate that it is inclined to the plane of the sky at an angle of 40 deg (previously it was estimated at 35 deg).

The two emission blobs seen in the low-excitation line [\ion{N}{ii}] 658\,nm (Fig. \ref{IFUoverplot} middle panel) indicate that the ionization front is positioned just at that location. To obtain an ionization front at a specified position there should be enough material between this front and the ionizing source. The axial symmetry of the observed feature guided our modelling towards a narrow and dense bar-like structure. Sufficient for a rather good fit was the assumption of a constant density with a value adopted to produce the ionization front at both far ends. Within this setup the middle region of this bar reproduces the \ion{He}{ii} 468\,nm unresolved central emission which requires intense ionization flux. Were the [\ion{N}{ii}] 658\,nm lines formed in two separate distant blobs the explanation of the unresolved \ion{He}{ii} 468\,nm emission would need a more elaborated nebular structure.

To reproduce the asymmetric H$\alpha$ image we attached  a plume to the bar-like structure, defined as a quarter of a circle of fixed thickness and with gas density decreasing away from the symmetry axis. This structure has the same kinematics as the bar-like outflows, therefore it is natural to place it near the axis and perpendicularly to the equatorial plane. The asymmetric plume-like structure extends to $\pm 2.4\times 10^{16}$\,cm above and below the plane of the image shown in Fig.\,\ref{density}.

To complete the model we added an outer equatorial low density torus.  

The total nebular mass of the proposed model is 0.09\,M$_\sun$; the central dense bar has a mass of 0.006\,M$_\sun$ while the plume-like feature has a mass of 0.003\,M$_\sun$. The chemical composition adopted for the photoionization modelling was taken from \citet{2009A&A...494..591C} and supplemented by the latest data from \citet{2017MNRAS.471.4648V}. The line ratios used for verification of the modelling were taken from \citet{2009A&A...500.1089G}.

In Table\,\ref{parameters} we compare the new model parameters with our previous analyses from 2014 and 2016 and with selected, representative observational data adopted from publications. We did not perform any automated search for a best-fit model such as the genetic algorithm used previously with spherically symmetric computations \citep{2006A&A...451..925G}, as the model is now too complicated for that. We started with the previously found basic parameters and searched for an improvement (with trials and errors, not far from the starting values), all the time verifying the agreement with the observed data. The nebular H$\beta$ flux was similarly well fitted with the older simpler codes, however details of line ratios are significantly better reproduced with the new 3D model. This in particular concerns the strong [\ion{O}{iii}] 500\,nm emission which is now excited by the higher temperature of the central star. Previously, lower value of $T_\mathrm{bb}$ was a compromise to fit high-excitation [\ion{O}{iii}] 500\,nm simultaneously with low-excitation {[\ion{N}{ii}]} 658\,nm for spherical symmetry. This conflict is now resolved by the spatial distribution of matter within the PN.

\begin{table}
\caption{Parameters of the models of PN H\,2-18 obtained with different codes compared with observed values. The line intensities are on the scale H$\beta$=1}
\label{parameters}
\centering
\begin{tabular}{l c c c c}
\hline\hline
   & \multicolumn{3}{c}{m o d e l l e d} & observed\tablefootmark{a} \\
                           & 2014\tablefootmark{b} & 2016\tablefootmark{c} &  2024 & \\
\hline
model version              & sphere     & ellipsoid &  3D        &        \\
ion. mass [M$_\sun$]      & 0.27       & 0.051     &  0.096     &        \\
kin. age [k\,yr]        & 1.67      & 1.28       & 2.4        &       \\
$T_\mathrm{bb}$ [kK]       & 51         & 52        &  62        &        \\
$L/L_\sun$                & 67\,000      & 1\,700     &  5\,000     &        \\
$\log F(\mathrm{H}\beta$)  & $-11.5$      & $-11.6$     &  $-11.5$     &  $-11.5$ \\
\ion{He}{ii} 468 nm        & 0.078      &           &  0.058     &  0.05  \\
{[\ion{N}{ii}]} 658 nm    & 0.0026     & 0.07      &  0.097     &  0.078 \\
{[\ion{O}{i}]}  630 nm     & 5$\times 10^{-8}$   & 8$\times 10^{-5}$  &  0.002 &  0.01  \\
{[\ion{O}{ii}]} 372 nm     & 0.005       & 0.07      &  0.13      &  0.16  \\
{[\ion{O}{iii}]} 500 nm     & 7.8        & 5.24      & 10.6       &  13.1  \\
{[\ion{Ne}{iii}]} 386 nm    & 0.70       &           &  0.57      &  0.999 \\
{[\ion{S}{ii}]} 673 nm      & 5$\times 10^{-6}$       &           &  0.013     &  0.018 \\
{[\ion{S}{iii}} 631 nm     & 0.0002    &           &  0.009     &  0.01  \\
{[\ion{Cl}{iii}]} 551 nm    & 0.007      &           &  0.081     &  0.0036\\
{[\ion{Ar}{iv}]} 471 nm     & 0.047      &           &  0.031     &  0.027 \\
\hline
\end{tabular}
\tablefoottext{a}{dereddened line intensities from \citet{2009A&A...500.1089G}}
\tablefoottext{b}{from \citet{2014A&A...566A..48G}}
\tablefoottext{c}{from \citet{2016A&A...585A..69G}}
\end{table}

With the adopted parameters the resulting best fitting model produces the image in the H$\alpha$ line in the plane of the sky presented in the right panel of Fig.\,\ref{mapHa} compared with the HST image in the left panel. Both images are rendered as an inverted colour map to better expose the faint structures and with contours added to guide the eye. It is not perfectly fitted and  it is not  unambiguous however it explains all the different observed characteristics and is as simple as possible.

\subsection{Velocity field}
\label{velocityfield}

\begin{figure*}
   \resizebox{\hsize}{!}{\includegraphics{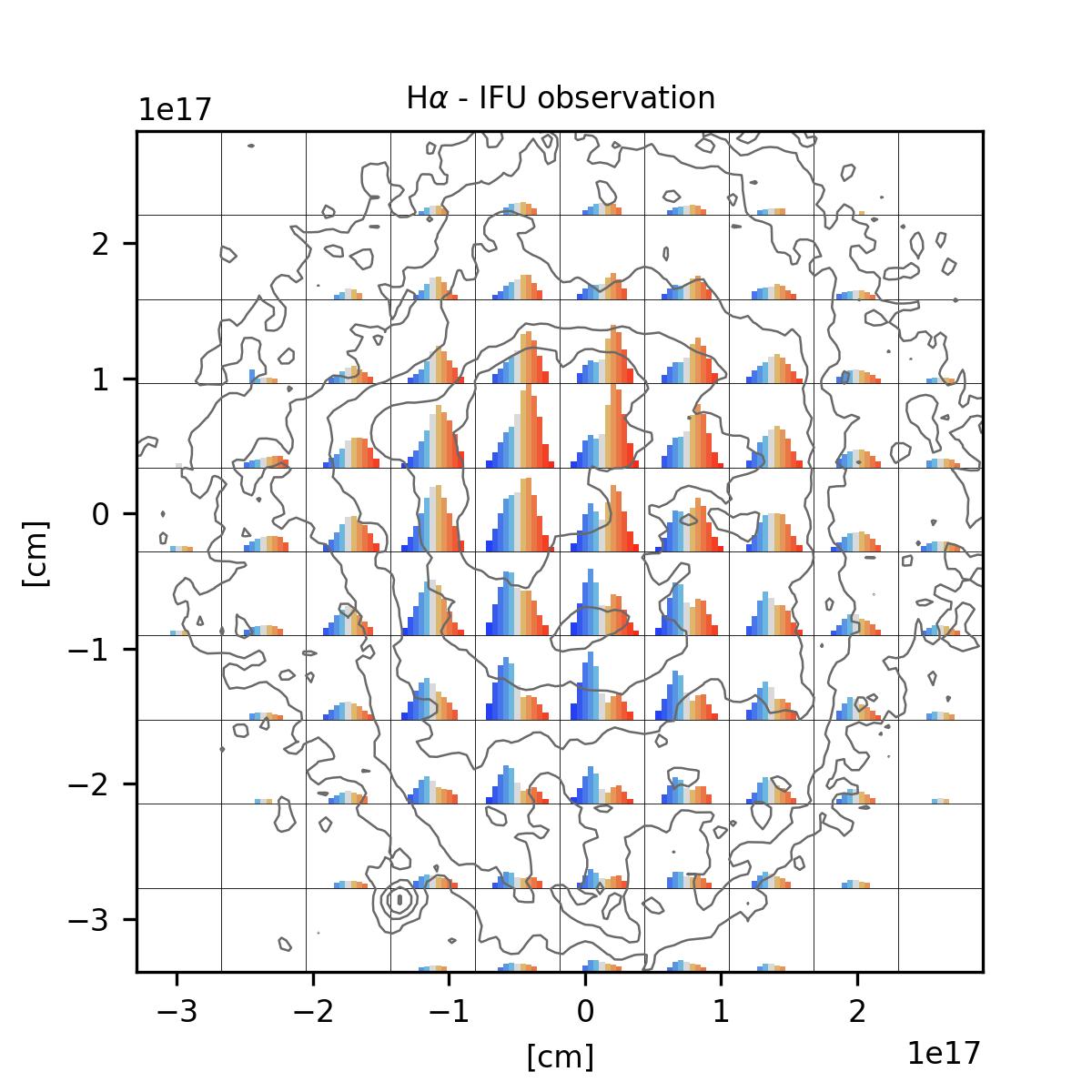}\includegraphics{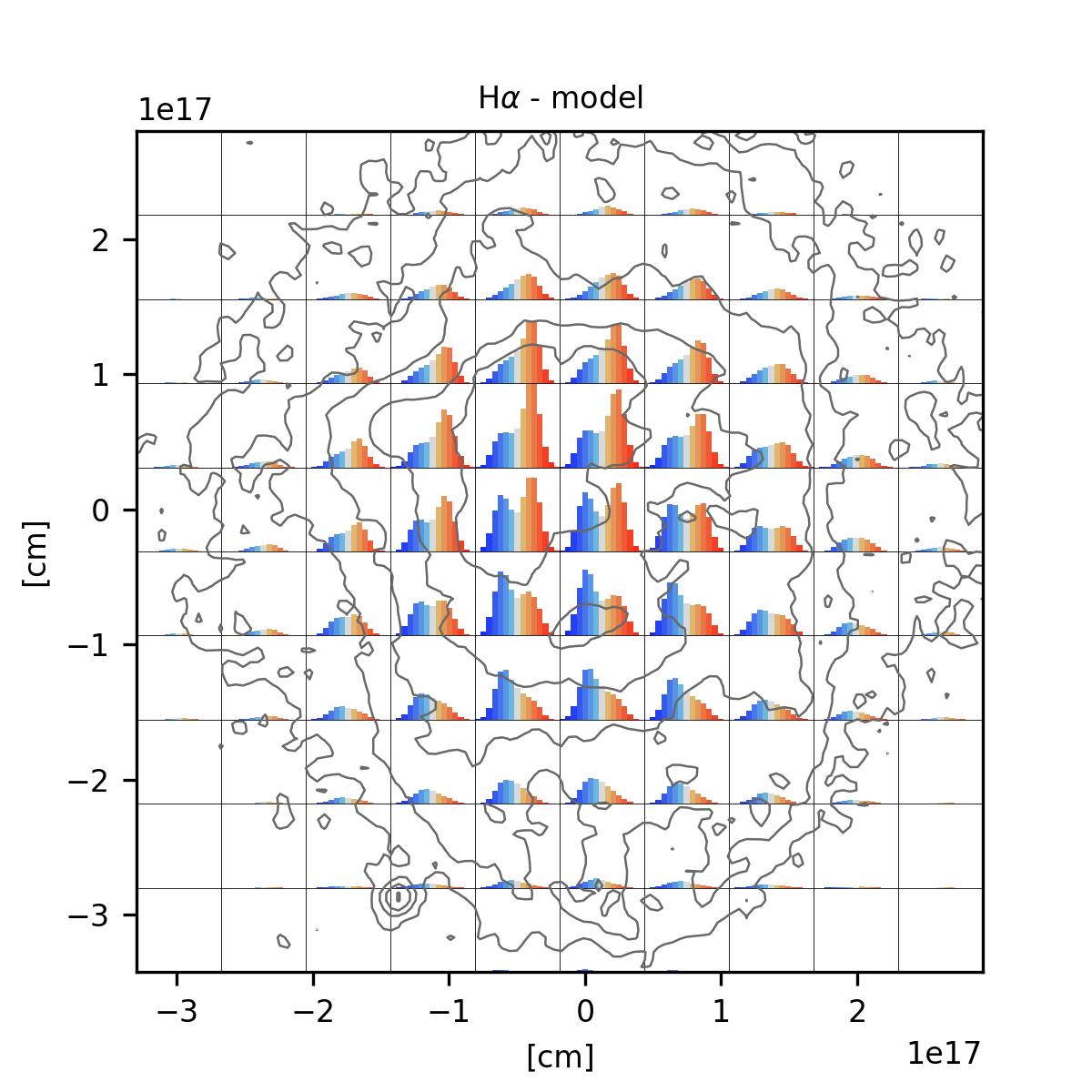}}
   \caption{The detailed emission profiles of the H$\alpha$ line shown for each of IFU pixels. The left panel shows the observed data, the right panel the corresponding calculated model emissions. To emphasize the value and direction of the velocity we applied  colour coding where red means gas moving away from the observer and blue moving towards the observer and the higher the velocity the more intense the colour, gray is for zero. We plotted the emission in a form of bar-plots, the width of each bar is 10\,km\,s$^{-1}$, the height (intensity) is normalized to the maximum value over the whole image. The horizontal size of each box extends at $\pm$75\,km\,s$^{-1}$, vertical from zero to unity.}
   \label{BarPloHa}
\end{figure*}

The ARGUS/IFU spectra allow much better for the velocity reconstruction than previously. However the abundance of data required some ingenuity to present clearly the 2D spectra and to compare observations with models. Fig.\,\ref{BarPloHa} shows the  presentation designed by us.

The left panel shows the observations, and the right panel the model results. The panels cover a fragment of the sky plane, the corresponding HST image (Fig.\,\ref{HSTima}) is overlaid as thin gray contours; the coordinates are expressed in units of $10^{17}$ cm.

Both panels are divided into $10 \times 10$ small squares representing the IFU pixels. Each square pixel shows the detailed emission profile of the H$\alpha$ line, with the horizontal axis expressed in km\,s$^{-1}$ over the range $\pm$75\,km\,s$^{-1}$ and the vertical axis extending from 0 to 1, normalized to the maximum value of the whole image. To emphasize the value and direction of the velocity we applied colour coding. As usual red means moving away from the observer (red-shifted) and blue moving towards the observer. The higher the velocity the more intense the colour; gray is for zero. We plotted the emission in the form of bar-plots to show better the colours. The width of each bar is 10\,km\,s$^{-1}$, the gray bar is centered on zero velocity.

In the observational panel on the left, we cut-off the noise at the 5\% level. In the model panel on the right we added a correction for the seeing. The seeing was below the requested 0.9 arcsec for most of the observations, except for the \ion{He}{ii} 468\,nm line. Inspecting the modelled emissions integrated over the exact pixel size it became obvious that we should allow for additional emission from some area around the given pixel which should mimic the actual seeing. We found a satisfactory fit (comparable to observations) when integration for the given pixel was extended by one pixel width (0.52 arcsec) around.

\begin{figure*}
   \resizebox{\hsize}{!}{\includegraphics{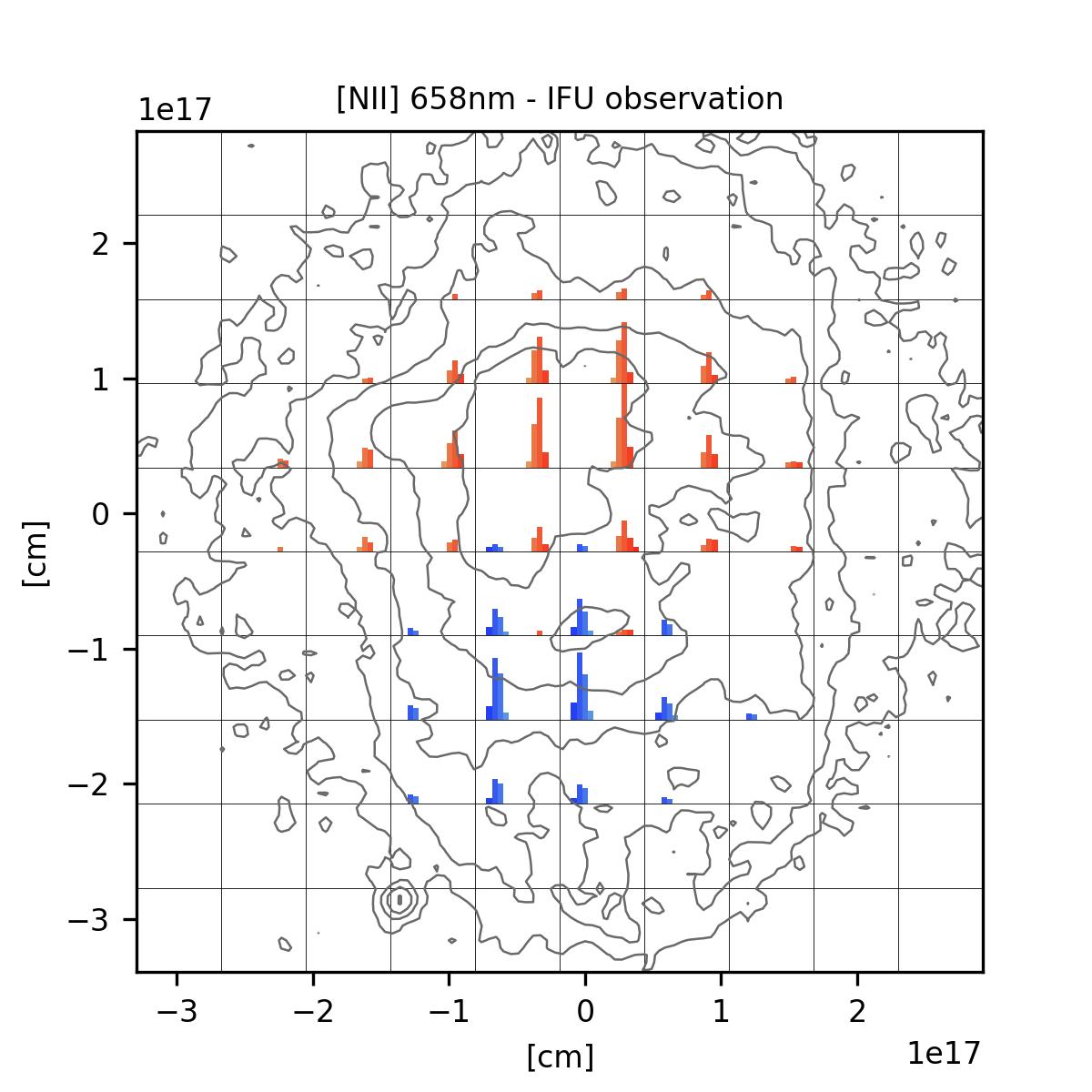}\includegraphics{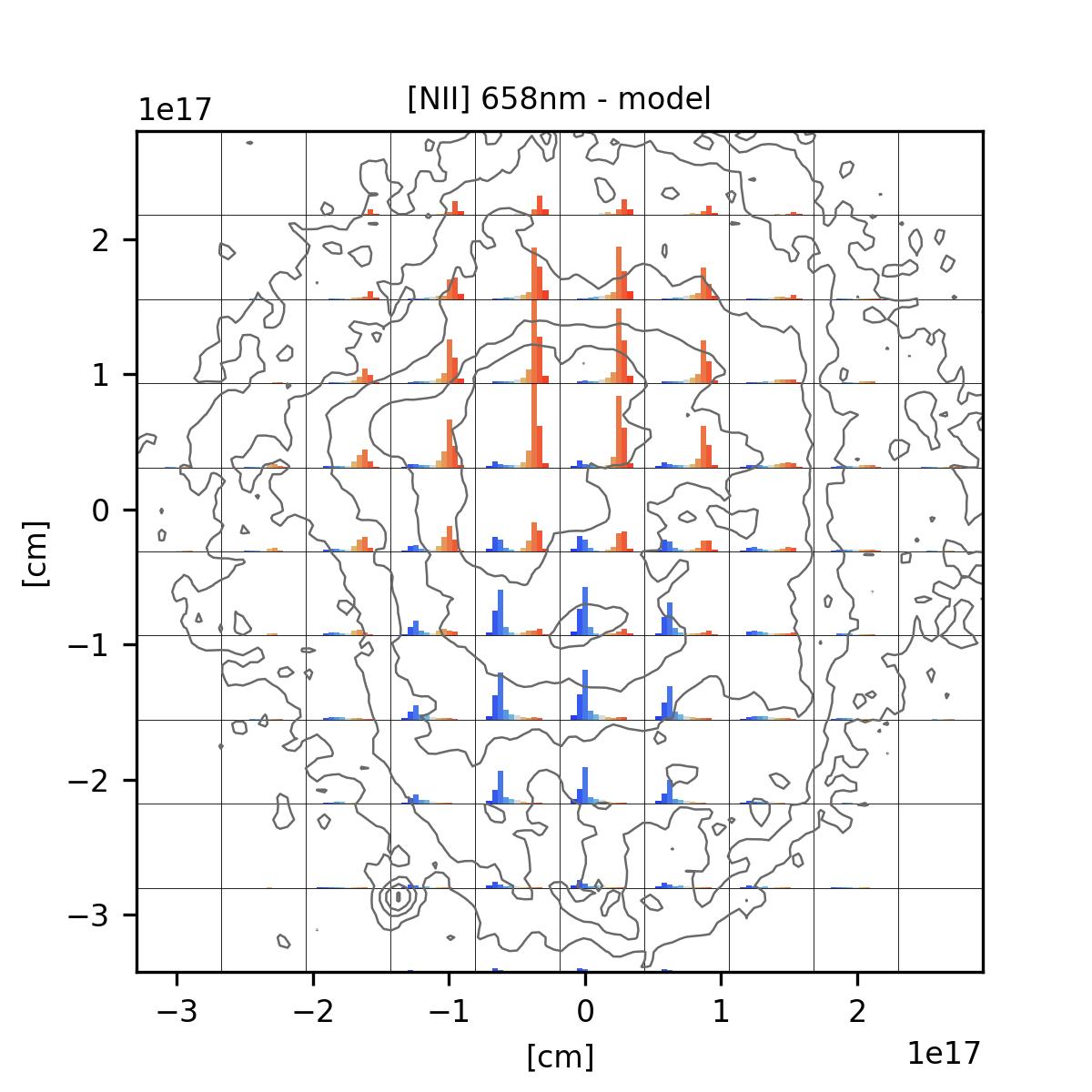}}
   \caption{The detailed emission profiles of the {[\ion{N}{ii}]} 658\,nm line shown for each of IFU pixels. The presentation is the same as in Fig.\,\ref{BarPloHa}}
   \label{BarPloN2}
\end{figure*}

In Fig.\,\ref{BarPloN2} the emission line {[\ion{N}{ii}]} 658\,nm is presented in the same setup as in Fig.\,\ref{BarPloHa}. Already at first glance we see that  red colour dominates in the upper half of the images, blue dominates in the lower half while the middle row is approximately symmetric. It is obvious that we see elongated structure and if we assume only expansion from the central source, then the upper part is directed away from us while the lower part is expanding  towards us. This also clarifies that the polar axis corresponds to the vertical direction in the image, something that is not obvious from the images only.

\begin{figure*}
   \resizebox{\hsize}{!}{\includegraphics{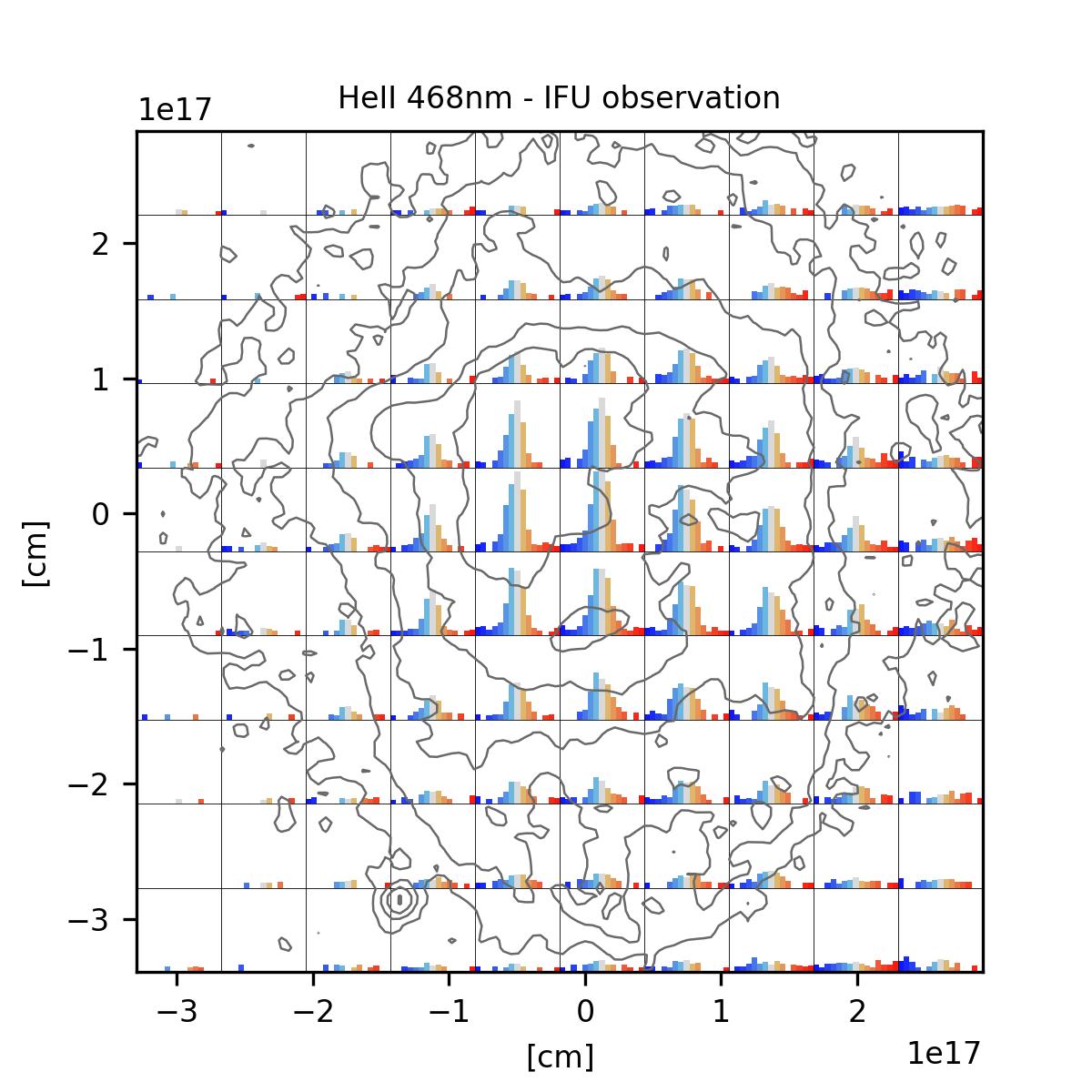}\includegraphics{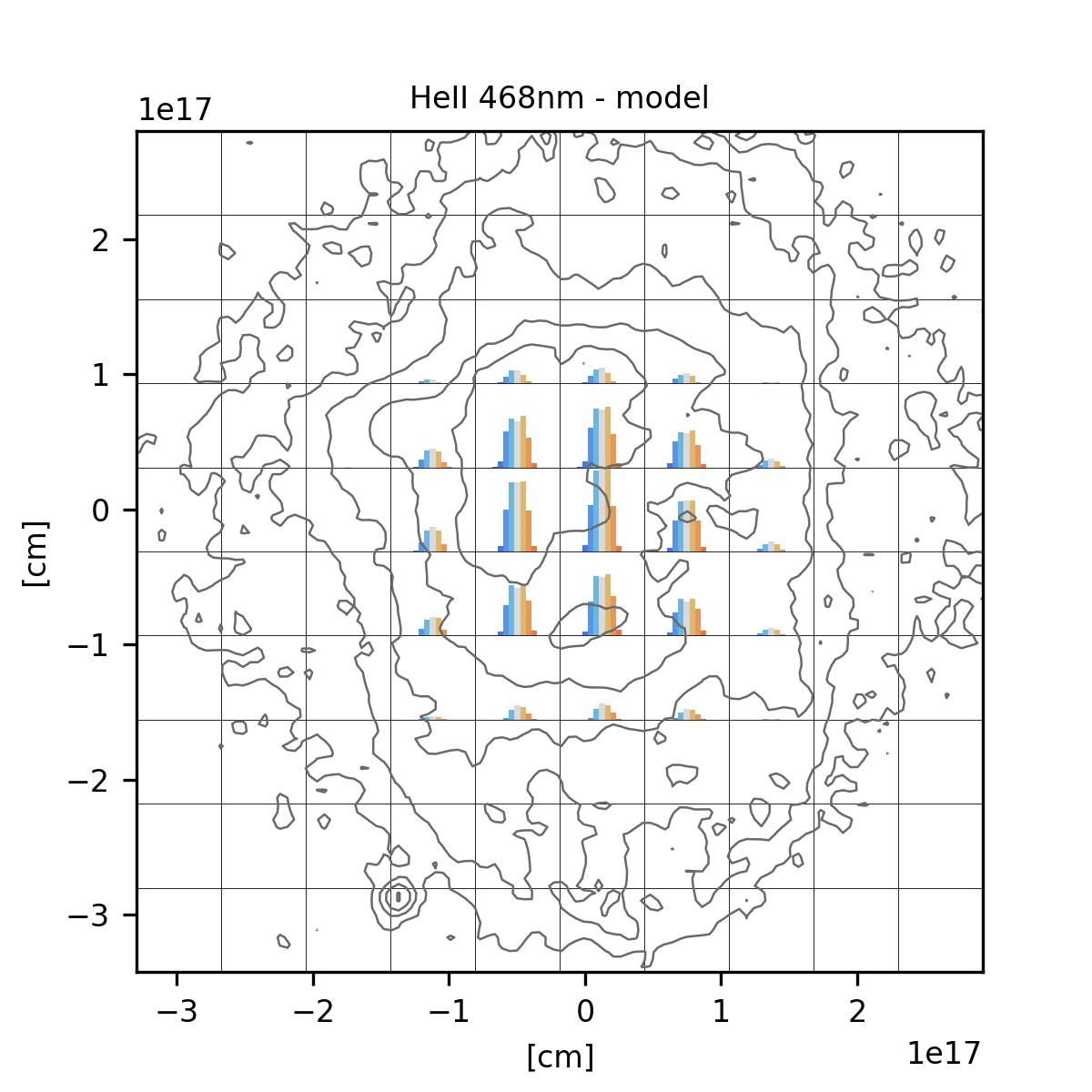}}
   \caption{The detailed emission profiles of the \ion{He}{ii} 468\,nm line shown for each of IFU pixels. The presentation is the same as in Fig.\,\ref{BarPloHa}}
   \label{BarPloHe2}
\end{figure*}

In Fig.\,\ref{BarPloHe2} the emission line \ion{He}{ii} 468\,nm is presented in the same setup as in Fig.\,\ref{BarPloHa}. Although of lower quality this line is unsplit and limited to the central pixels, in contrast to the [\ion{N}{ii}] emission.

The left panels of Figs.\,\ref{BarPloHa}--\ref{BarPloHe2} show clearly that no extraordinary kinematics is detected within H\,2-18. The velocities are within normal ranges for PNe, and there is no evidence for a fast-moving and/or collimated jet. 

The data allows us to improve on the velocity model of \citet{2016A&A...585A..69G} where the velocity was linearly increasing with radial distance. Now we have spectral coverage over the whole nebula. Interestingly the velocities of the different substructures identified previously, as seen in the H$\alpha$ line, are similar to each other. This suggests an isotropic velocity field. Indeed, attempts to apply different velocities to the different substructures failed to improve the fit, in agreement with this. (There may be differences at smaller scales: we are limited by the spatial resolution of the IFU data cube.)

A radial gradient of the expansion is obviously present because the high excitation \ion{He}{ii} 468\,nm line shows the lowest velocity while the low excitation {[\ion{N}{ii}]} 658\,nm line has the highest. A very simple relation of velocity linearly increasing with distance (homologous expansion) was not satisfactory because it makes the {[\ion{N}{ii}]} 658\,nm line too broad in comparison with observations. The region of the [\ion{N}{ii}] line formation should rather have a constant velocity. We adopted a velocity monotonically increasing with radial distance, starting from 10\,km\,s$^{-1}$ at the inner edge and reaching an asymptotic value of 60\,km\,s$^{-1}$ near the ends of the bar-like structure. In this way we obtained a remarkable fit to the broad, split, and asymmetrical H$\alpha$ emission and simultaneously to the narrow, fast and oppositely directed \ion{N}{ii} emissions and spatially unresolved, narrow, unsplit \ion{He}{ii} 468 nm line, as indicated by the high spectral resolution of the IFU data.

%________________________________________________________________

\section{Discussion}
\label{discussion}

\subsection{New details and improved parameters}

The collected IFU spectra provided new information about the Galactic Bulge nebula H\,2-18. Although of lower spatial resolution than HST they revealed a feature proposed to be a bar-like structure embedded inside the main object and co-axial with it. We call it a `bar' rather than `jet' because its relatively low velocity is similar to that of the  surrounding broader cylinder. The adjoining single-sided plume of ejected gas inclined to the main axis was earlier seen in the HST images. Its kinematics is also not different from the rest of the nebula. The ears-like extensions seen in the HST image are now simply explained as part of the outer equatorial torus.

The photoionization modelling in 3D allowed for improved estimation of stellar and nebular parameters. The higher $T_\mathrm{bb}$ now better reproduces the line ratios, the advantage of 3D density structure is obvious in this context. The $T_\mathrm{bb}$ value indicates a not very old object, likely in the middle of its way towards White Dwarf cooling track. The size of this PN agrees with that: the kinematic age (derived from mass averaged velocity of 43\,km\,s$^{-1}$, lobes lengths of 0.15\,pc, and the corrected formula of \citet{2014A&A...566A..48G}) is 2400\,yr.

\subsection{Co-axial structures}

\begin{figure}
   \resizebox{\hsize}{!}{\includegraphics{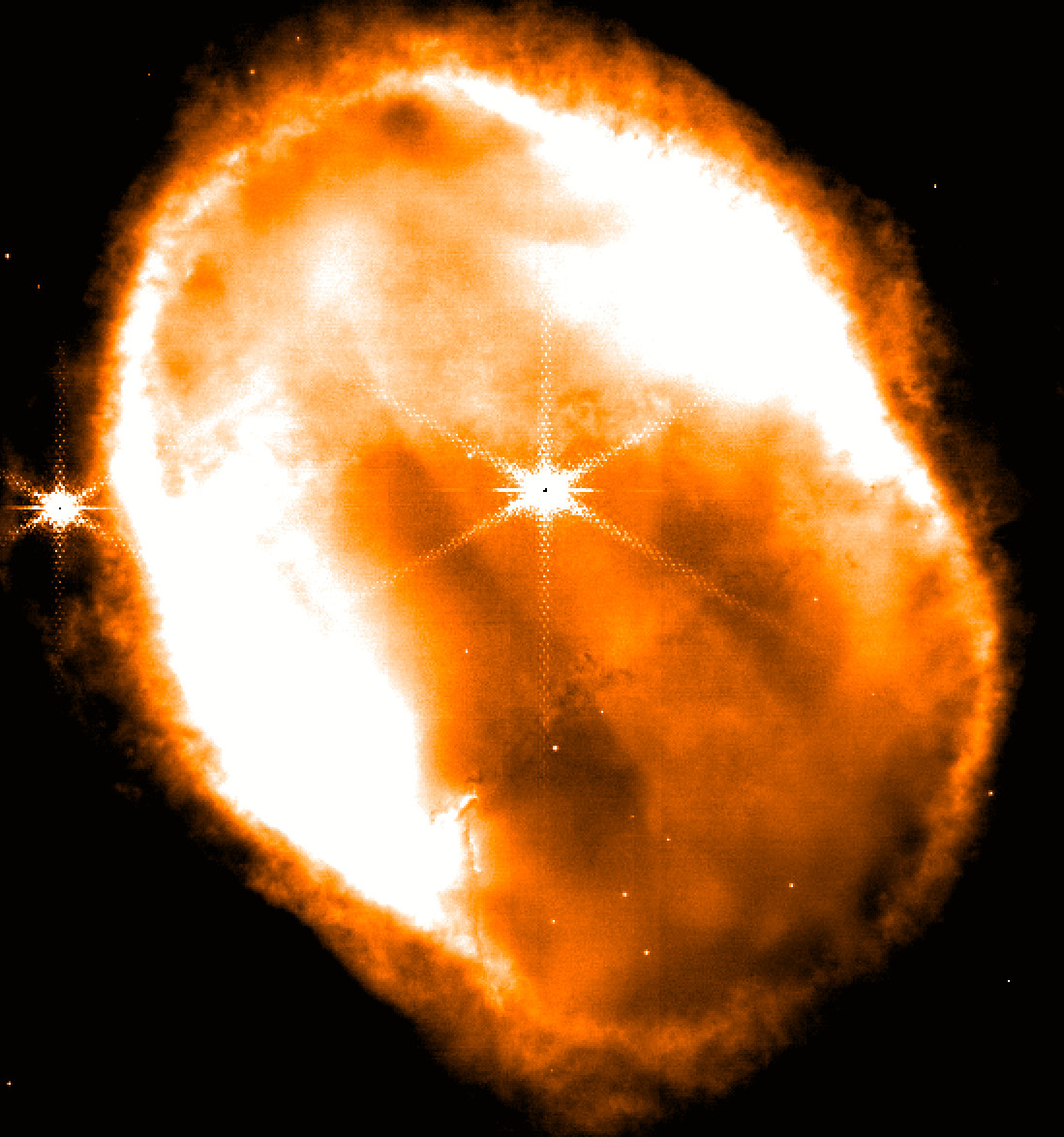}}
   \caption{JWST image of NGC\,3132 in the filter NIRcam F187N dominated by Pa$\alpha$ emission, showing a bar-like, polar structure with some similarity to that in H\,2-18. Image from \citet{2022NatAs...6.1421D}.}
   \label{N3132}
\end{figure}

\citet{2024A&A...684A.107M} analysed the integral field spectroscopy of the PN A\,58 and derived a structure of wide bipolar outflows with an embedded narrow co-axial collimated stream. Their kinematics (bipolar outflows at 280\,km\,s$^{-1}$) is different from ours but the morphology looks nearly the same (certainly excluding the plume). A common envelope influence is suggested. Their analyses were performed with the well known tool named SHAPE\footnote{https://wsteffen75.wixsite.com/website} which is focused on morphology and does not consider the photoionization. A thin bar-like structure in the polar direction is also seen in the JWST image of the planetary nebula \object{NGC 3132} \citep{2022NatAs...6.1421D}, shown in Fig. \ref{N3132}. The equatorial ring is seen in absorption against the bar, which indicates that the bar is located inside the nebula. Both cases show that the co-axial inner and outer nebula in H\,2-18 is not a unique phenomenon. (A thin bar-like feature  in JWST images of NGC\,6720 \citep{2024MNRAS.528.3392W} is a different structure, as it is mainly seen as a gap and is located in the equatorial direction: it is interpreted as  the inner edges of a wide bipolar flow.)

Comparison  with published hydrodynamical calculations is hampered by the fact that  very few of them  show the common envelope evolution outcome on a planetary-nebula wide scale and are rendered as images (most of them are limited to the solar radius or a.u. scale and they are usually presenting intricately shaped density cross-sections). \citet{2018ApJ...860...19G} present a common-envelope model that is extended into the PN phase. It assumes a pre-existing equatorial density enhancement into which the common envelope ejecta expand. It is restricted to axial symmetry with numerous simplification, and is focused on explaining bipolar shapes while skipping kinematics. Their Fig.\,8 shows two examples of emission measures (a proxy for images) for complicated elongated nebulae tilted by 40 deg which show structures similar to those in H\,2-18:  the outer barrel shape, central ellipsoidal enhancements, and symmetric extensions (ears). One of their models shows blobs of gas piled-up at the symmetry axis which look similar to the symmetric \ion{N}{ii} emissions found by us. Interestingly two separate mass ejection events are not required for the formation of both inner and outer lobes. A follow-up study \citep{2022MNRAS.517.3822G} concluded that lower mass stars will produce more elongated PNe  and the observed jets in PNe must be remnants of early phases. Both conclusions might be applicable to our case.

\subsection{Low velocity outflow}

The spherical models of \citet{2014A&A...566A..48G} commonly revealed a velocity increase towards the outer nebular edge. This is in agreement with spherically symmetric hydrodynamical modelling of \citet{2005A&A...431..963S}. Here we obtained a much more complex 3D structure, still with a similar velocity field. We can speculate that the same acceleration mechanism known from hydrodynamical models is acting here: the slowly spreading gas is accelerated and heated by the increasing ionizing radiation of the gradually heating-up central star.

One of the conclusions of \citet{2018ApJ...860...19G} was that the inner nebulae can be ablated and photoevaporated from the excretion disk. One can expect that such a process can result in a mild acceleration of the PN.  A disk wind, caused by the evaporation of the photoionized gas was our interpretation of low velocity outflows in \object{PN M 2-29} \citep{2010A&A...514A..54G}. Recently \citet{2022Galax..10...53I} analysed a circumstellar ring irradiated by the central star and obtained that star-driven evaporation would produce a cylindrically collimated outflow. The simple model was intended to reproduce the very elongated PN Hen\,3-401. Interestingly the cylindrical outflow appears to be relatively slow (below 10\,km\,s$^{-1}$ near the irradiated disk) and presumably weakly accelerated along the axis. In our Fig.\,\ref{HSTima} we do not see any obvious disk comparable to the one seen in Hen\,3-401, nevertheless some similarities exist. Our model of H\,2-18 (shown in Fig.\,\ref{density}) has lengths comparable to Hen\,3-401  and approximately twice the width \citep{2022Galax..10...53I}. The structure requires higher density in the equatorial region which can be treated as an approximation to a true equatorial disk or an equatorial density enhancement. Therefore, photo-evaporation from this region with further photo-acceleration are a plausible mechanism shaping H\,2-18. 

These models are not perfect matches for H\,2-18: the lobes in \citet{2022Galax..10...53I} have low mass, while for the models of \citet{2018ApJ...860...19G} most of the common envelope mass does not reach escape velocities and remains bound. The latter model also does not present the kinematics.

Circumbinary disks in PNe have low mass \citep[e.g.,][]{2022NatAs...6.1421D} and may have difficulty explaining the dense cylindrical structure in H\,2-18. A larger-sized equatorial density enhancement may however play a similar role.

\subsection{Jets or not}

\citet{2021ApJ...913...91A} propose that structures such as the `ears' in H\,2-18 form through polar jets, while   \citet{2018ApJ...860...19G} form them by equatorial gravitational focusing.  In the case of H\,2-18, the `ears' are presumably located in the equatorial plane so that the second mechanism is more likely.

The cylindrical polar structure is more ambiguous. In \citet{2018ApJ...860...19G}, it forms through expansion  of the common envelope ejecta into a pre-existing equator-to-pole density gradient. \citet{2022Galax..10...53I} shows that the cylindrical nebula of Hen\,3-401 can originate from an evaporating disk (or torus?), without requiring a jet.  In contrast, \citet{2002ApJ...568..726S} shows that cylindrical structures in PNe can form through refocused jets. The paper uses a jet velocity of 500\,km\,s$^{-1}$, as could come from a main sequence companion. Such velocities are not seen in H\,2-18, nor is a jet bow shock seen, but the structure could have formed from a previous jet which is now extinct. 

\subsection{Binary central star}

There is no  direct evidence of a close binary star in H\,2-18 centre, nevertheless the axial symmetry of the dominant structures suggests the possibility of a binary system shaping this PN. All the discussed hydrodynamical models require binarity. However, none appear to depend on common-envelope evolution.  Wider binary interaction which avoids common envelope can still produce equatorial density enhancement, which may suffice. A binary (multiple?) system with a disk can be well hidden in the central unresolved dense region.

\subsection{Asymmetric plume}

The one-sided plume-like structure can be compared with the one observed in the PN M\,2-29 \citep{2010A&A...514A..54G}, however in H\,2-18 it expands a little faster, it is seen at different angle, and is located inside the main nebula although protruding a little out of it. Its flat structure with a nearly constant density was proposed on  the basis of the rather uniform intensity seen in H$\alpha$ image; any plowing action of the ionization front is not seen. The mass and brightness of the plume are comparable to those of the bar which makes it difficult to disentangle both components. This simple model does not reproduce all details so the true shape may be more complicated. The models of \citet{2018ApJ...860...19G} show structures with a rough similarity to the plume, although with point-symmetric multiplicity that is lacking in H\,2-18.

The mass loss shaping of H\,2-18 has in any case been a complex process. The high density of the inner bar or cylinder, the plume and the lack of a detected jet put significant constraints on its evolution.

%__________________________________________________________________

\begin{acknowledgements}
We thank Vincent Icke the referee for friendly and helpful comments.
Help from Roger Wesson during installing and running the MOCASSIN code is gratefully acknowledged. We acknowledge financial support from the Nicolaus Copernicus University through the University Centre of Excellence `Astrophysics and astrochemistry'. Part of this work was supported by STFC through grant ST/X001229/1. AAZ also acknowledges support from the Royal Society through grant IES/R3/233287 and the University of Macquarie. This work made use of Astropy:\footnote{http://www.astropy.org} a community-developed core Python package and an ecosystem of tools and resources for astronomy \citep{2022ApJ...935..167A}.

\end{acknowledgements}

\bibliographystyle{aa}
\bibliography{H_2-18_AandA.bib}

\end{document}